# New Prototype Multi-gap Resistive Plate Chambers with Long Strips

Y.J. Sun[a], C. Li[a], M. Shao[a], B. Gui[a], Y.E. Zhao[a], H.F. Chen[a], Z.B. Xu[b], L.J. Ruan[b], G.J. Lin[c], X. Wang[d], Y. Wang[d], Z.B. Tang[a], G. Eppley[e], P. Fachini[b], M. Kohl[f], J. Liu[e], W.J. Llope[e], R. Majka[c], T. Nussbaun[e], E. Ramberg[h], T. Sakuma[f], F. Simon[i], N. Smirnov[c], B. Surrow[f], D. Underwood[g]

[a] *Department of Modern Physics, University of Science and Technology of China, Hefei 230026, China*

[b] *Brookhaven National Laboratory, Upton, NY 11973, USA*

[c] *Yale University, New Haven, Connecticut 06520, USA*

[d] *Department of Engineering Physics, Tsinghua University, Beijing, 100084, China*

[e] *Rice University, Houston, Texas 77251, USA*

[f] *Massachusetts Institute of Technology, Cambridge, MA 02139-4307, USA*

[g] *Argonne National Laboratory, Argonne, Illinois 60439, USA*

[h] *Fermi National Accelerator Laboratory, Batavia, Illinois 60510, USA*

[i] *Max-Planck-Institute for Physics and Excellence Cluster "Universe", Munich, Germany*

**Abstract** A new kind of Multi-gap Resistive Plate Chamber (MRPC) has been built for the large-area Muon Telescope Detector (MTD) for the STAR experiment at RHIC. These long read-out strip MRPCs (LMRPCs) have an active area of 87.0×17.0 cm$^2$ and ten 250 μm-thick gas gaps arranged as a double stack. Each read-out strip is 2.5 cm wide and 90 cm long. The signals are read-out at both ends of each strip. Cosmic ray tests indicate a time resolution of ~70 ps and a detection efficiency of greater than 95%. Beam tests performed at T963 at Fermilab indicate a time resolution of 60-70 ps and a spatial resolution of ~1 cm along the strip direction.

## 1 Introduction

Multi-gap Resistive Plate Chambers (MRPC) [1, 2] are inexpensive gaseous detectors with good timing resolution and high detection efficiency. The readout pads can be shaped into arbitrary shapes to achieve the granularity required in modern heavy-ion collisions. This technology has been chosen for the STAR and PHENIX Time-of-Flight (TOF) systems[3, 4] at RHIC, the ALICE TOF system[5] at the LHC, to name just a few. If the charged particle multiplicities in the events are not too large, MRPCs with long double-ended read-out strips (LMRPCs) are attractive as these can reduce the number of electronics channels and also provide a measurement of the hit position along each strip via "left minus right" timing.

This LMRPC technology has been selected for the proposed large-area Muon Telescope Detector (MTD) [6] for the STAR experiment. The MTD has been proposed to measure muons of momenta of a few GeV/*c*. Such a system will allow the detection of di-muon pairs from Quark Gluon Plasma (QGP) thermal radiation, quarkonia, and light vector mesons. The correlation of quarks and gluons as QGP resonances, Drell-Yan production, and the measurement of heavy flavor hadrons via semi-leptonic decays into single muons, are also possible. These measurements will advance our knowledge of the nuclear matter formed in the relativistic heavy ion collisions at RHIC.[7] Two LMRPCs were constructed, and have been installed in STAR as a prototype MTD detector. They have successfully taken data since the 2007 run of RHIC. To meet the physics requirements of the MTD, the LMRPCs are required to have a time resolution that is better than 100 ps and a spatial resolution of each hit better than 2 cm. In this paper, the LMRPC prototypes and their performance using both cosmic rays and test beams are described.

## 2 LMRPC Prototype

The LMRPC module has ten gas gaps arranged in a double stack as shown schematically in Fig. 1. The thickness of each gap is 250 μm and is defined using nylon monofilament fishing line. Float glass sheets of 0.7 mm thick with volume resistivity of ~$10^{13}$ Ω·cm are used as the resistive plates. The positive and the negative high voltage (HV) is applied to adhesive graphite tape (with surface resistivity of ~200 kΩ/□) located outside the outermost glass of each stack. The total size of each LMRPC is 95.0×25.6 $cm^3$, and the active area defined by the read-out strips is 87.0×17.0 $cm^2$. The read-out pads are segmented into six double-ended strips which are 2.5 cm wide. The gap between each strip is 4 mm wide. Pins are used to transmit the negative signals from the external strip layers to the central strip layer. Twisted-pair cables bring the differential signals from both ends of the module to the front-end electronics. These front-end electronics preamplify and amplify the signals and drive the signals into 50 Ω over long cables to CAMAC ADCs (charge Analog-to-Digital Converter) and TDCs (Time-to-Digital Converter). The thickness of a finished LMRPC is about 3.2 cm with two pieces of 0.8 cm thick honeycomb boards supporting the whole structure. The weight of one module is around 7.2 kg. The LMRPC modules are enclosed in a gas-tight aluminum box that is flooded with a mixture of 95% Freon R-134a and 5% iso-butane.

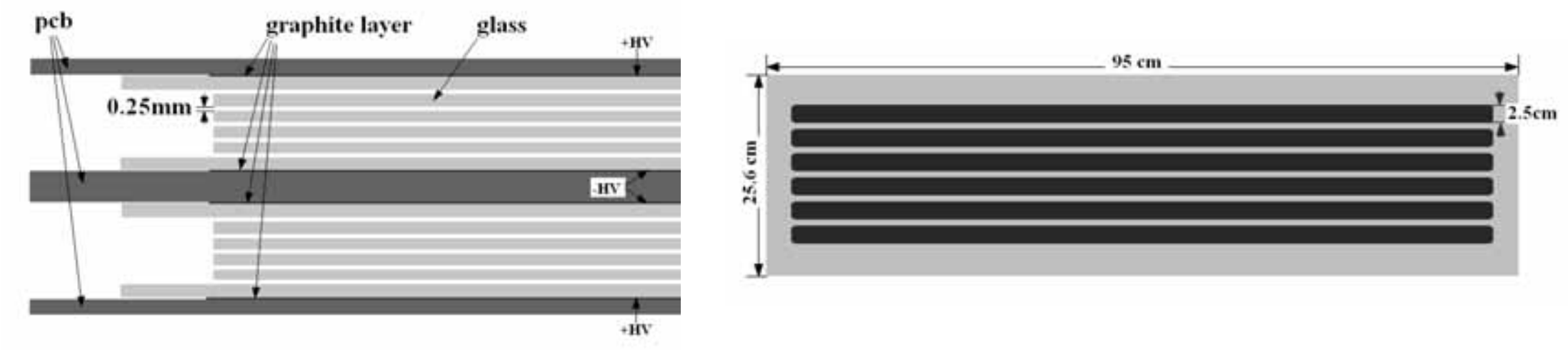

Fig. 1 Schematic side-view of the LMRPC modules (left frame) and read-out strips (right frame).

## 3 Cosmic ray test

The prototype module was tested with cosmic rays using the set-up shown in Fig. 2. Three scintillators were used to define a triggered area of 20×5 $cm^2$ along the read-out strips. This trigger area fully covered one LMRPC strip in width (the "strip under investigation") and half each of the two neighboring strips. Two of these scintillators were read-out at both ends with four fast PMTs which provided the event reference time (T0). A STAR-TOF MRPC module [8] also covered the triggered area with six readout pads, which divided the triggered area into six segments with area of 3×6 $cm^2$. The LMRPC signals were read out from both ends via the front-end electronics based on MAXIM 3760 chips. These electronics provided both amplification and discrimination. For additional details on the cosmic tests, see Ref. [9].

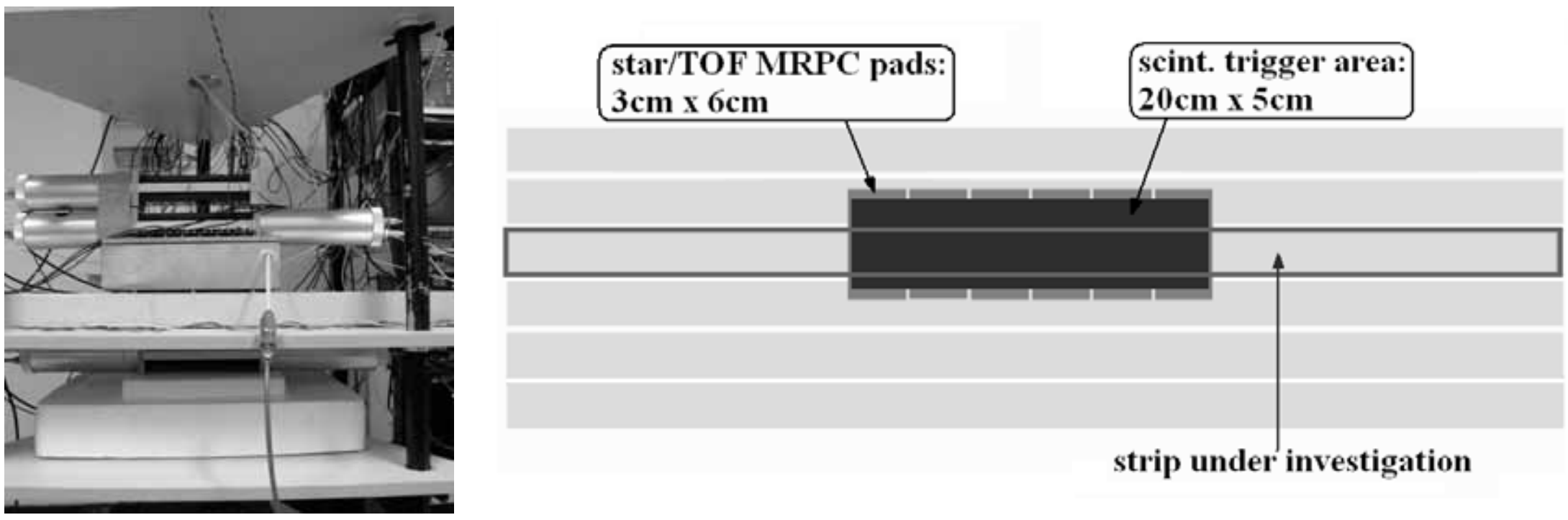

Fig. 2 The set-up used for the cosmic ray tests.

The plateau of detection efficiency versus HV is shown in Fig. 3. The detection efficiency is above 95% for positive and negative HV values greater than 6.2 kV. For our subsequent measurements, the positive and negative HV values used were 6.3 kV.

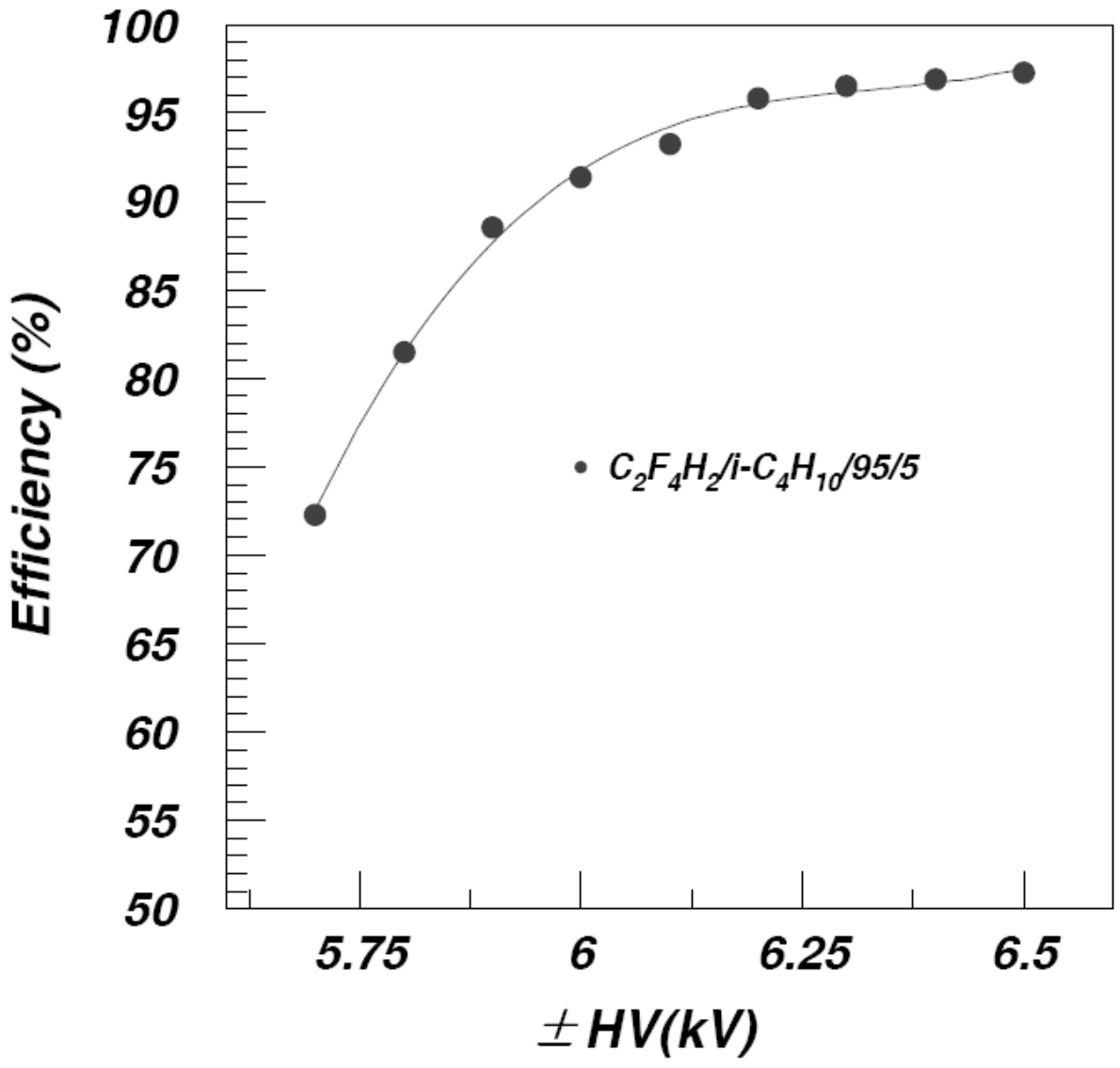

Fig. 3 LMRPC detection efficiency plateau.

The six pads of the STAR-TOF MRPC gave a rough position along the readout strips. Fig. 4 shows the relationship between one-half of the time difference measured from the two ends of the investigated strip and the relative hit position. The slope of the linear fit indicates the reciprocal of the signal propagation velocity along the strip, which is observed to be 59.6 ps/cm.

Fig. 5 shows the relationship between the charge of the signals taken from the two ends (left and right) of the read-out strip under investigation. The left frame of this figure shows this relationship when the triggered area was placed in the middle of an LMRPC strip, while the right frame shows this relationship when the triggered area is at an extreme end of the read-out strip. Even in the latter case when the signal transmitting distances to left and right are quite different, a linearly correlation can still be observed. This indicates that the attenuation of pulse area along the strip is neglectable.

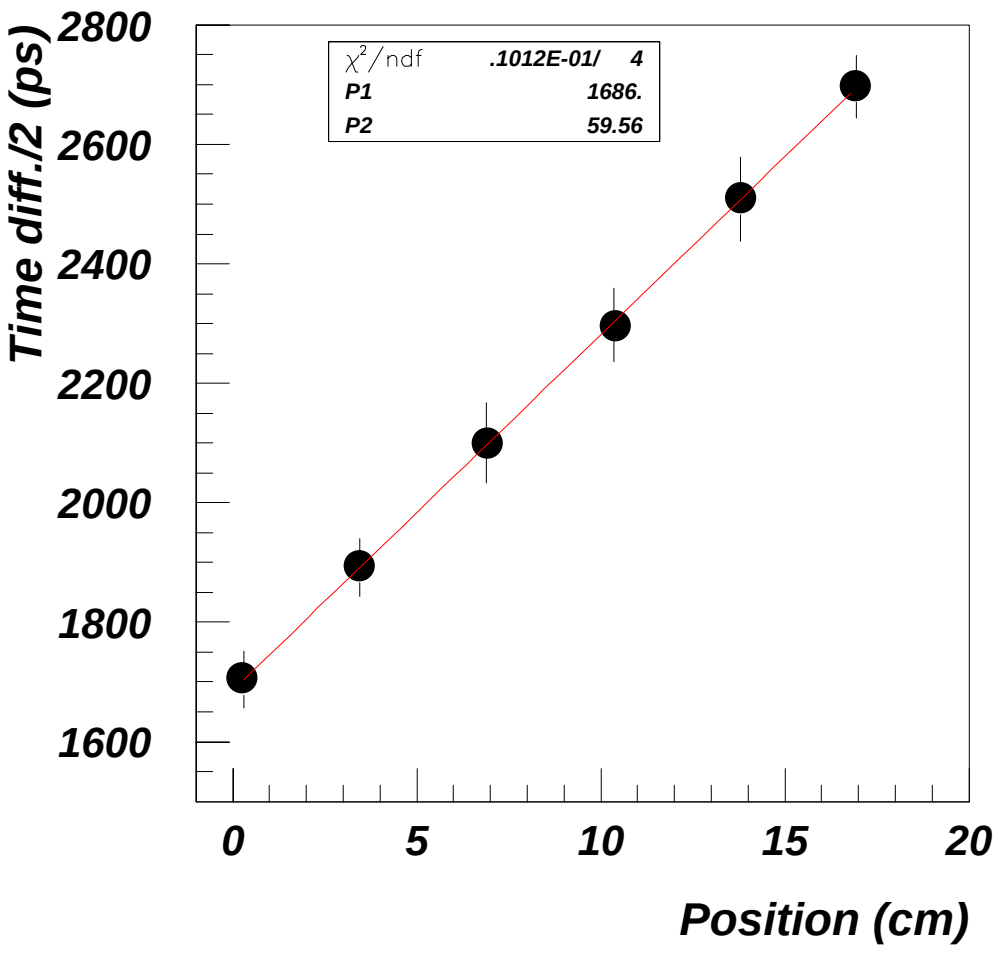

Fig. 4 One-half of the time difference measured from the two ends of the investigated strip versus the hit position. The signal propagation velocity along the LMRPC strip is the slope of this line.

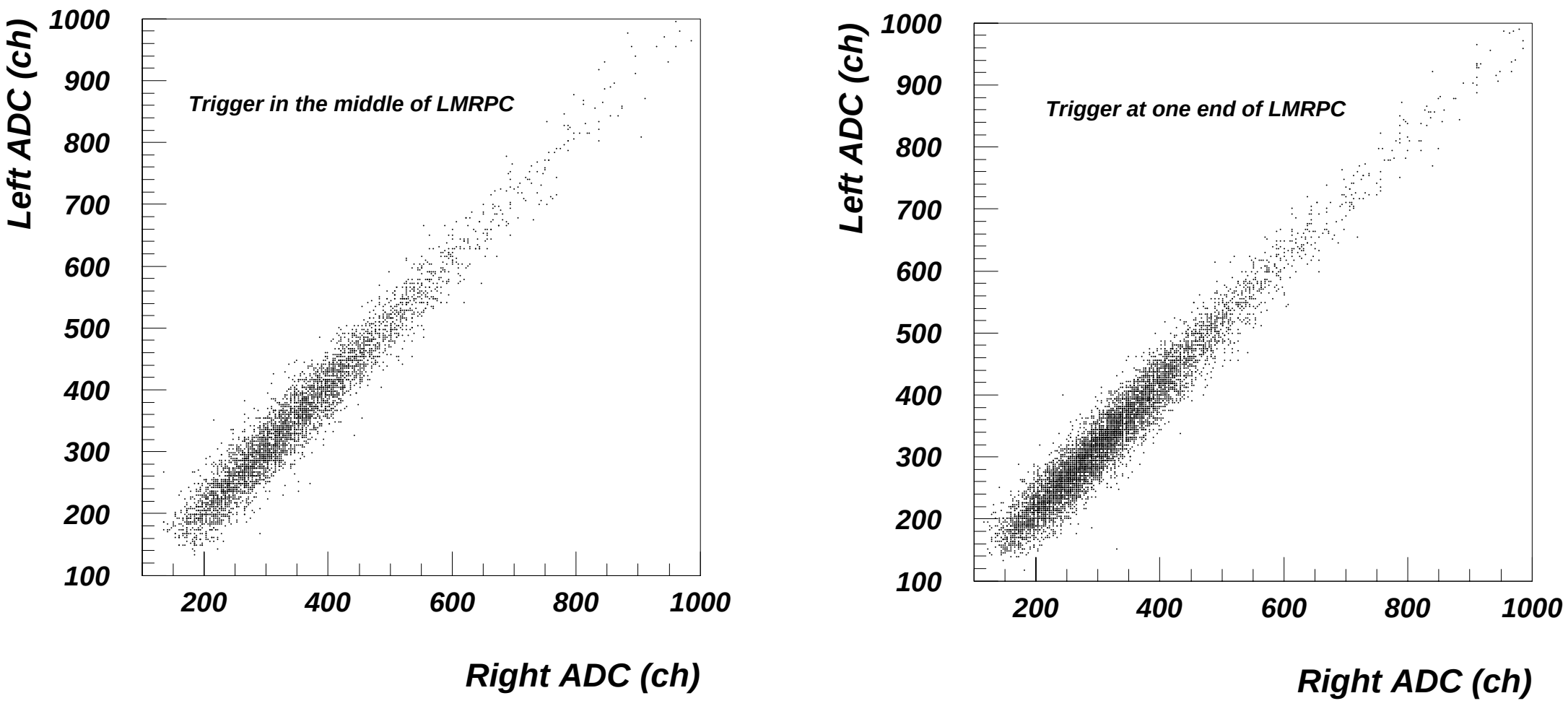

Fig. 5 The correlations of the signal areas from the two ends of one readout strip for hits near the mid-point (left frame) and one end (right frame).

The mean time, $T_{mean}=(T_{left}+T_{right})/2$, formed from the time from the two ends of one strip eliminates the time smearing caused by the position of the hit along the strip. The signals from two ends had been found to be similar in area. Thus, the slewing correction was made between the arithmetical mean of time and pulse area (mT-mA correction), and is shown in Fig. 6a. A sixth-order polynomial was fit to these curves to obtain the slewing correction function. The reference time resolution was 98 ps, as shown in Fig. 6b. The distributions of mean time with respect to the reference time after the mT-mA correction are shown in Fig. 7. Following the subtraction of the reference time resolution in quadrature, the time resolution of the LMRPC itself was observed to be 60 ps (74 ps) when triggered in the middle (at one end) of the module.

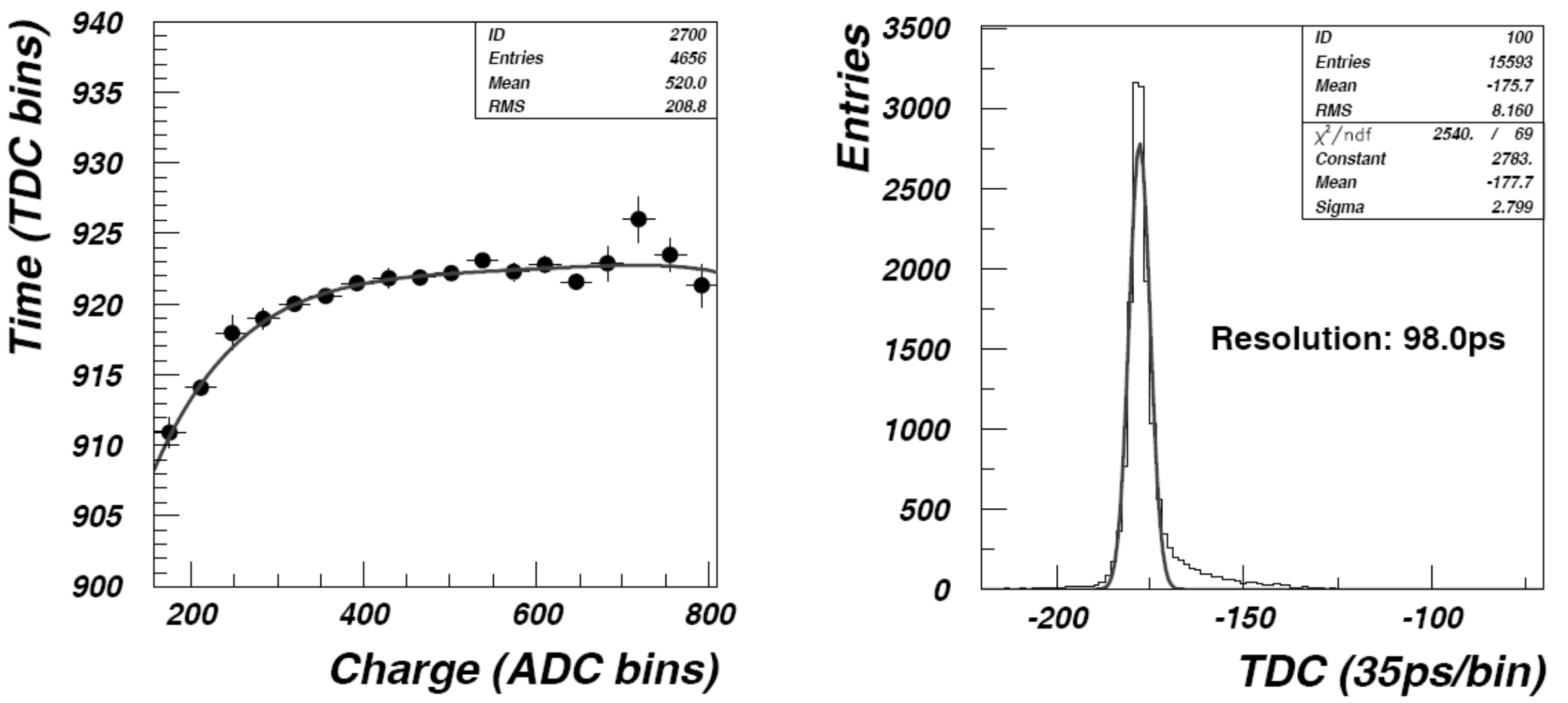

Fig. 6 The mT-mA slewing correction (left frame) and the resolution of the reference time, T0 (right frame).

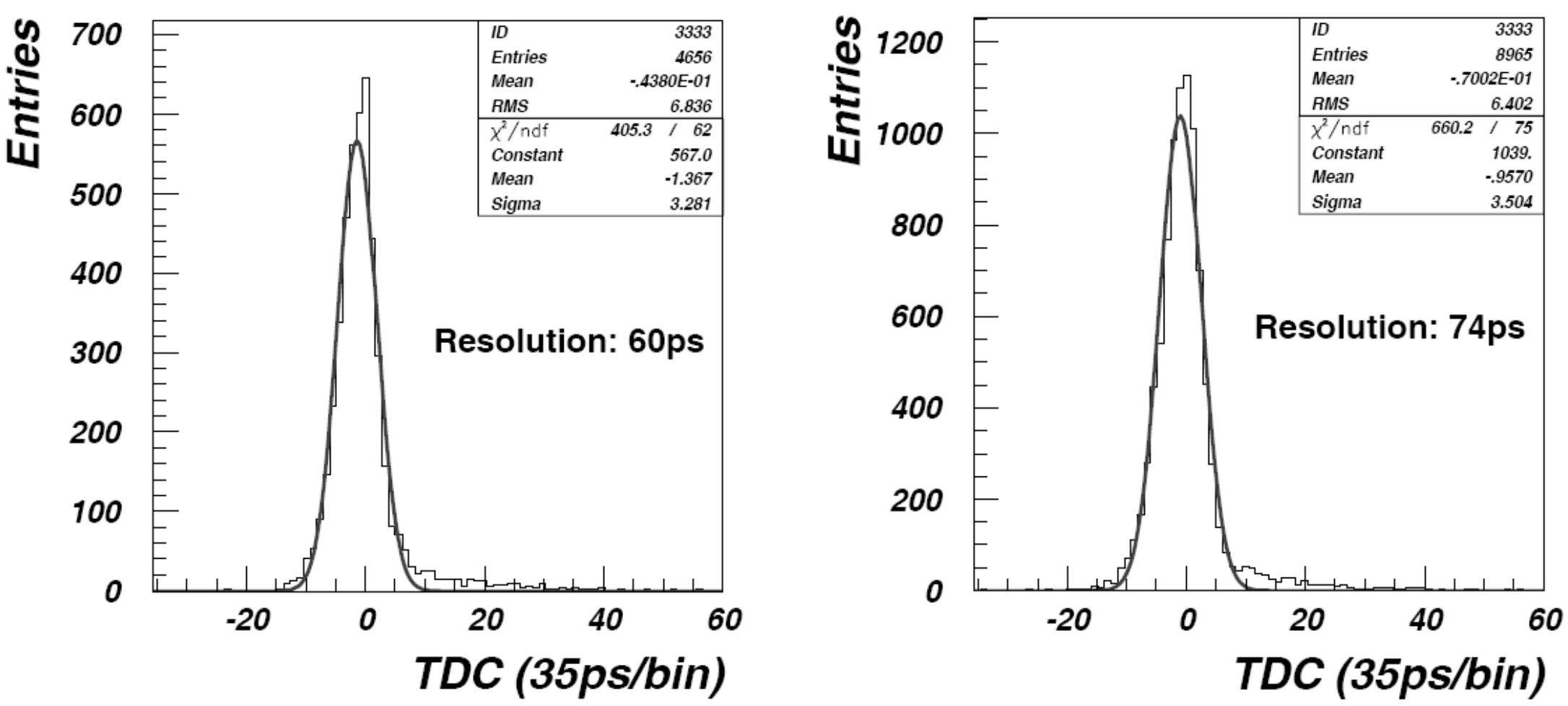

Fig. 7 Mean time resolution: (left) trigger in the middle and (right) trigger at one end of a strip.

## 4 Beam test at FNAL

A beam test experiment (T963) was carried out at the MTEST facility at Fermi National Accelerator Laboratory (FNAL) in May 2007. The beam consisted primarily of 32 GeV muons, pions, and protons. The set-up of the beam test is shown in Fig. 8. Two LMRPC modules were tested. One was made at Tsinghua University (MRPC1, upstream) and the other was made at the University of Science and Technology of China (MRPC2, downstream). These were placed on a movable platform. TOF1, TOF2 and TOF3 are three layers of scintillators. The coincidence of TOF1 and TOF2 was used as the event trigger and the common start signal for the TDCs. TOF3 provided the reference time for the LMRPCs. Nine layers of Multi-Wire Proportional Chambers (MWPCs) and three layers of Gas Electron Multiplier (GEM) chambers provided the beam particle position at the LMRPCs. Two Cherenkov detectors (C1 and C2) were used for beam particle identification.

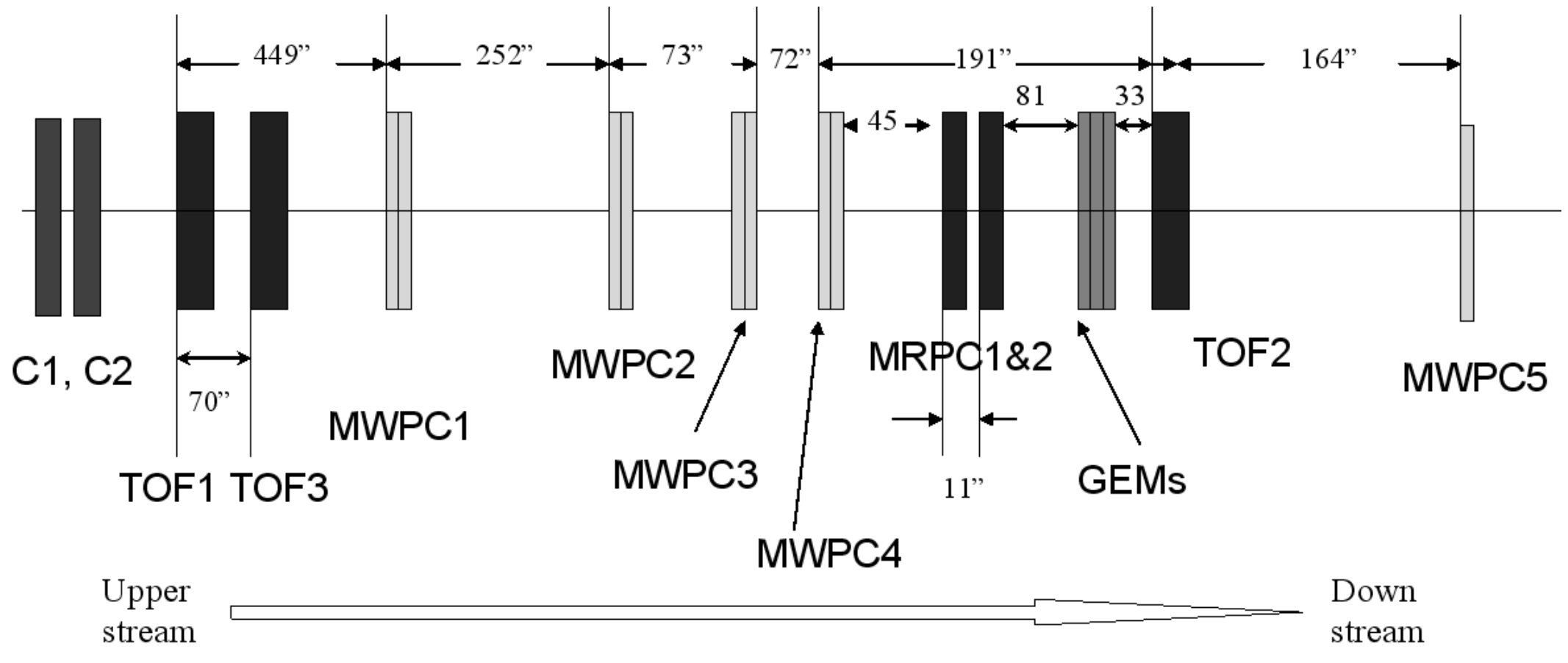

Fig. 8 T963 beam test set-up.

Fig. 9 shows the detection efficiency plateau of the USTC module. The electric field E in the gas gaps is equal to the applied HV values (the absolute sum of the positive and negative values) divided by the total thickness of all of the gas gaps (250 μm × 5) in one stack. Over a wide range of working electric field values (90-104 kV/cm), the detection efficiency is greater than 98%.

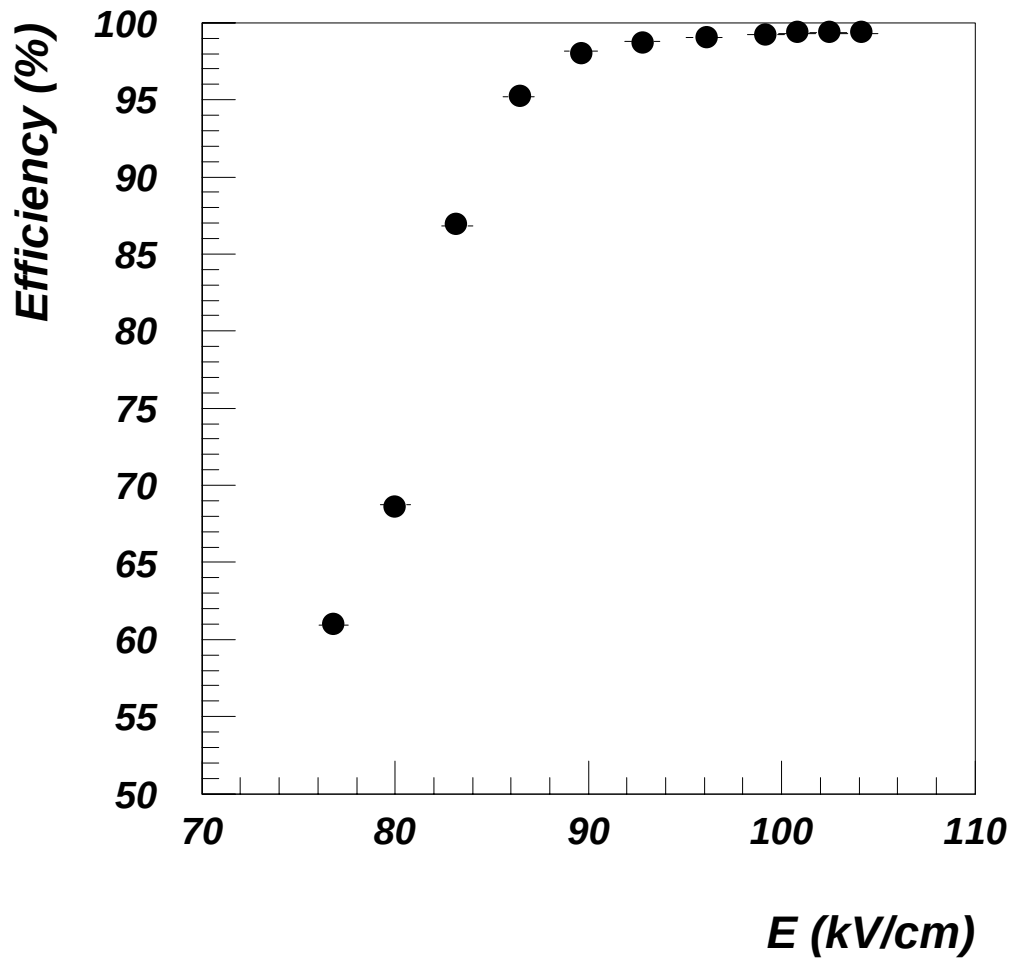

Fig. 9 Detection efficiency plateau from the beam test.

After the slewing correction and the subtraction in quadrature of the reference time resolution, the LMRPC time resolution is plotted versus the electric field and shown in Fig. 10. The time resolution is less than 70 ps for electric fields between 100-105 kV/cm.

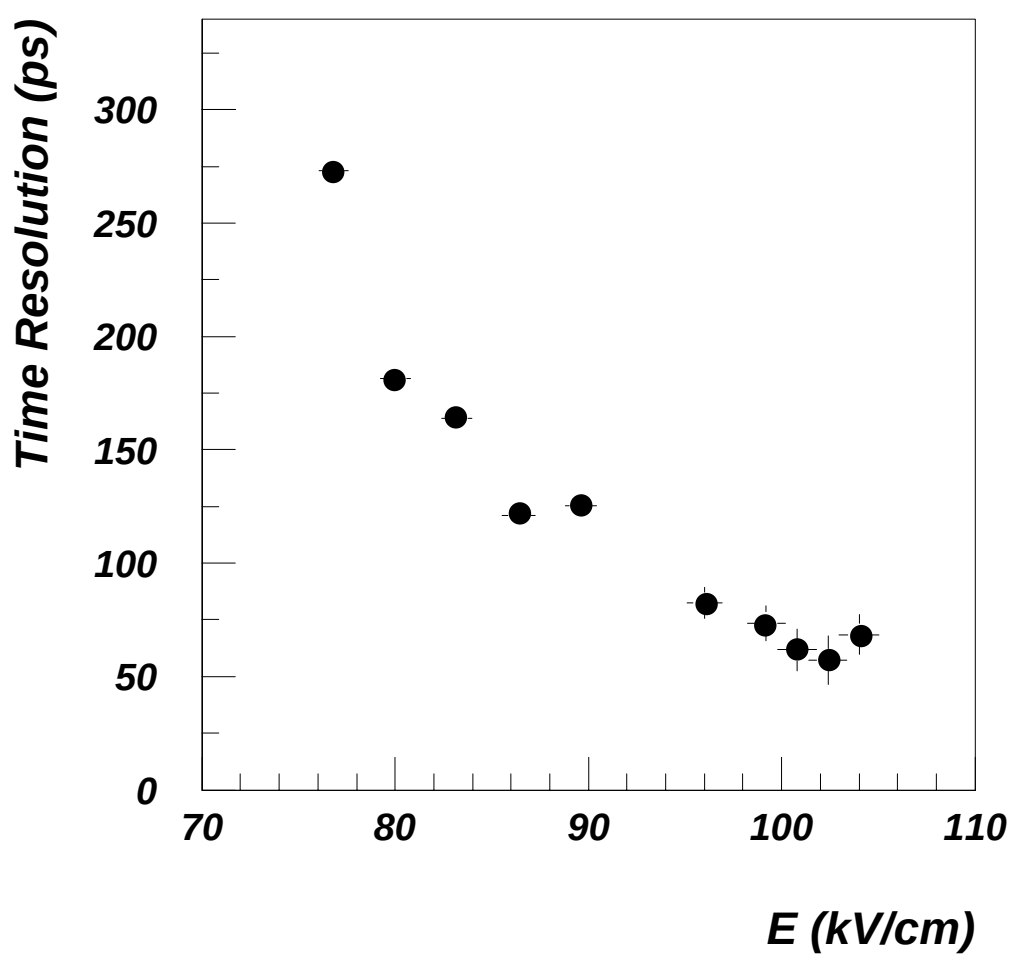

Fig.10 Time resolution as a function of electric field.

The correlation between the difference of the times measured from the two ends of one LMRPC strip and the particle incident position is shown in Fig. 11. The strong correlation that is observed confirms that the time information can be used to calculate the position of the hit along the strip. The slope of the linear fit indicates a signal propagation velocity of ~60 ps/cm, which is consistent with that obtained from the cosmic ray tests. Shown in Fig 12a is a scatter-plot of the beam particle position as measured by the GEM detectors versus the LMRPC hit position obtained with the difference of the times at the two ends of one strip. The variance around this correlation, shown in Fig. 12b, is assumed to result purely from the LMRPC spatial resolution and is observed to be of 0.6 to 1.0 cm since the spatial resolution of GEM is found to be around 70 μm [10].

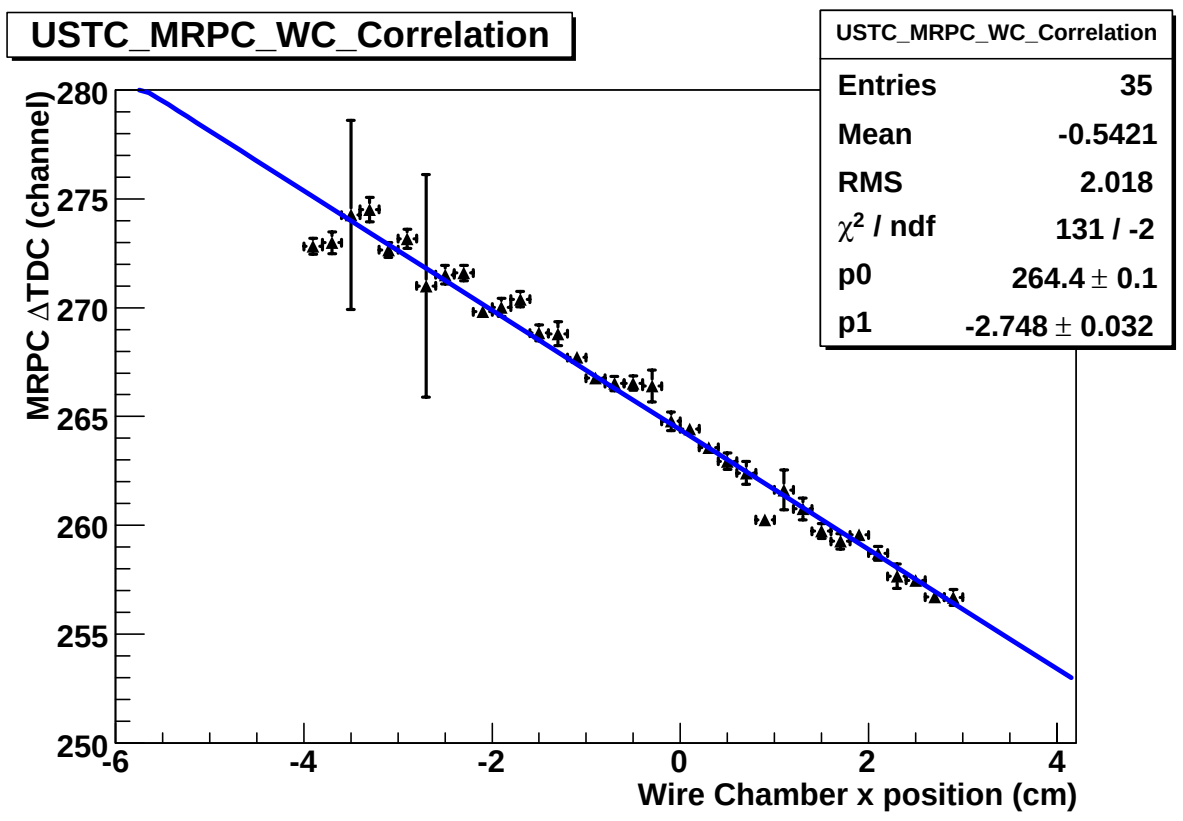

Fig.11 Correlation between the LMRPC time difference and charged particle position measured by MWPC.

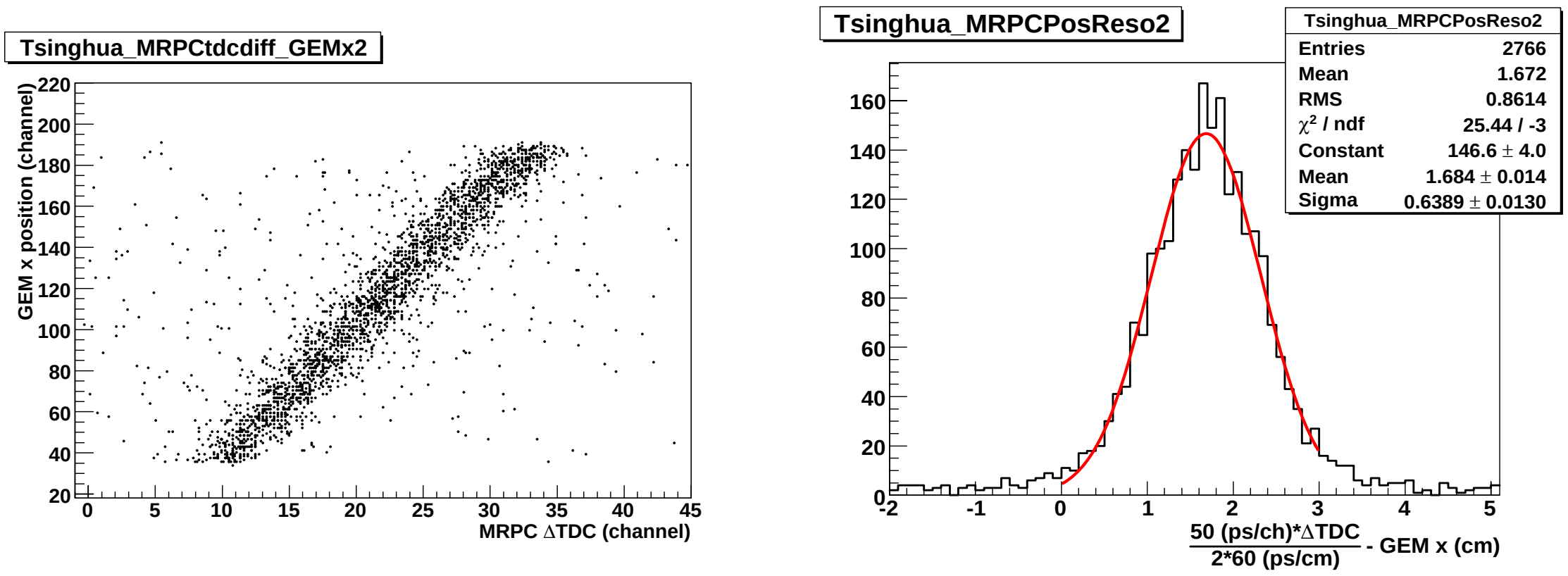

Fig. 12 a) The hit position from the GEM chambers versus the hit position from the LMRPC time differences and b) spatial resolution of LMRPC.

The beam energy was also tuned down to 4 GeV. The Cherenkov detectors were used to separate protons from pions and muons. At this beam energy and in the present set-up, the time-of-flight of the protons from TOF1 to LMRPC (~ 890 inches, as indicated in Fig. 8) should be 2.5 ns (50 channels) longer than that for the pions and muons. These times of flight are shown in Fig 13. The protons are well-separated from the pions and muons.

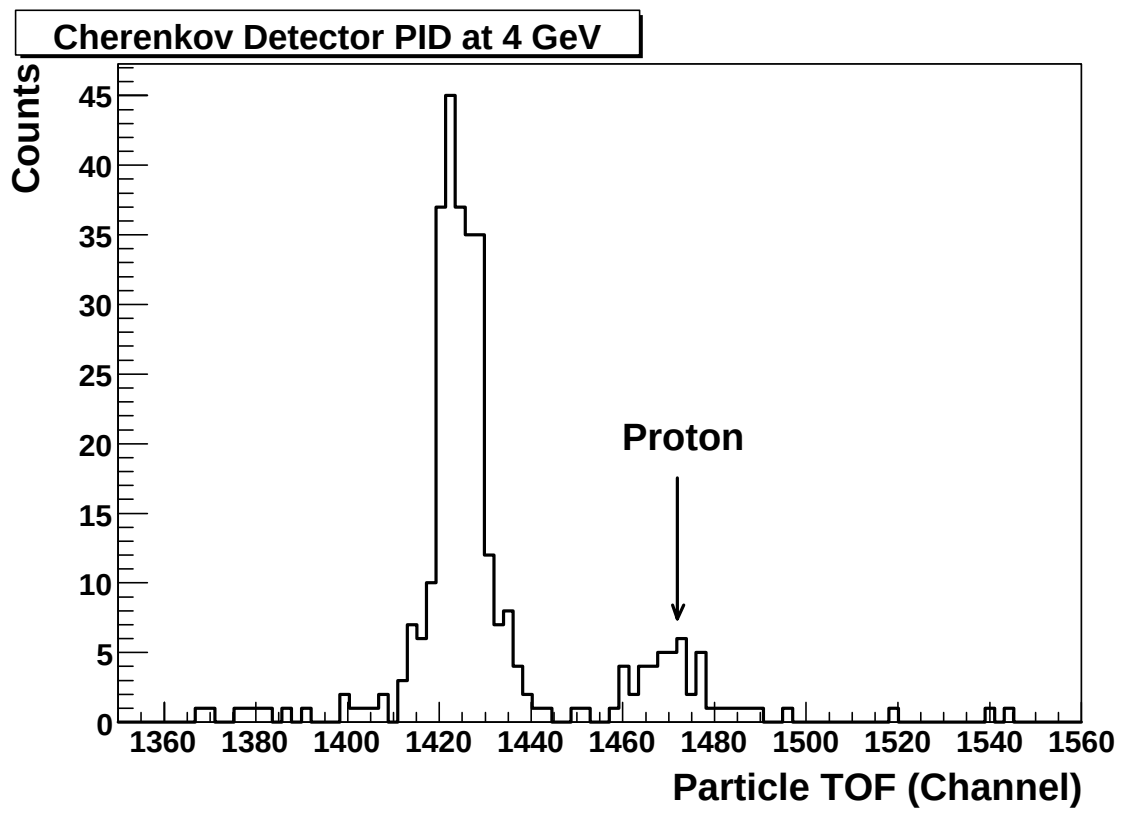

Fig. 13 Proton identification by time of flight measurement of LMRPC.

## 5 Summary

The new prototype LMRPCs have been built and tested both with cosmic rays and test beams. The cosmic ray tests indicate a time resolution of ~70 ps and a detection efficiency greater than 95% at a detector high voltage of ± 6.3 kV. In the T963 beam test performed at FNAL, similar values of the detection efficiency and the time resolution were observed. The spatial resolution of these LMRPCs along the direction of the read-out strips was near or less than 1 cm. The performance of these LMRPC prototypes thus satisfies the requirements for the STAR MTD system, and also appears applicable for Time-of-Flight at the electron ion collider eRHIC [11].

## Acknowledgements

This work is supported by National Natural Science Foundation of China (10775131, in part

10620120287, 10610285, 10675072 and 10775082), Knowledge Innovation Program of the Chinese Academy of Science, BNL LDRD project (07007) and China Postdoctoral Science Foundation (20070410784). We thank Fermilab for the allocation of beam time and the support of beam operations. One of us (Lijuan Ruan) would like to thank the Battelle Memorial Institute and Stony Brook University for support in the form of the Gertrude and Maurice Goldhaber Distinguished Fellowship.

## References

[1] M.C.S. Williams, E. Cerron, et al., Nucl. Instr. and Meth. A434 (1999) 362.

[2] CHEN Hong-Fang, LI Cheng, et al., HEP&NP, V26 N3 (2002) 201 (in Chinese).

[3] B. Bonner, H. Chen, G. Eppley, F. Geurts, J. Lamas-Valverde, Ch. Li, W.J. Llope, T. Nussbaum, E. Platner, and J. Roberts, Nucl. Inst. and Meth. A508 (2003) 181.

[4] Proposal for a Large-Area Time-of-Flight system for STAR, STAR-TOF Collaboration, June 2003, http://wjllope.rice.edu/~TOF/TOF/Documents/TOF-20040524.pdf.

[5] ALICE Time of Flight Addendum, CERN/LHCC 2002-016, Addendum to ALICE TDR 8, 24 April 2002.

[6] "A New Large-area Muon Telescope Detector at Mid-rapidity at RHIC", 2006 Division of Nuclear Physics Annual Meeting, October 25–28, 2006, Nashville, Tennessee, USA.

[7] Zhangbu Xu, "A novel and compact muon telescope detector for QCDLab", BNL LDRD project.

[8] SUN Yong-Jie, LI Cheng, et al., Nucl.Sci. and Tech. V16 N4 (2005) 231.

[9] HUANG Sheng-Li, LI Cheng, XU Zi-Zong, et al., HEP&NP V27 N2 (2003) 154.

[10] Frank Simon, James Kelsey, Michael Kohl, Richard Majka, Miroslav Plesko, David Underwood, Tai Sakuma, Nikolai Smirnov, Harold Spinka, Bernd Surrow, arXiv:0711.3751.

[11] RHIC-II/eRHIC white paper (Dec. 2006), http://www.bnl.gov/npp/docs/RHICplanning/rhic2_wp_draft_dec30_2006.pdf